\documentclass{midl} 

\usepackage{color,soul}
\usepackage{multirow}
\usepackage{mwe} 

\title[\footnotesize{Does Anatomical Contextual Information Improve 3D U-Net-Based Brain Tumor Segmentation?}]{Does Anatomical Contextual Information Improve 3D U-Net-Based Brain Tumor Segmentation?}

\midlauthor{\Name{Iulian Emil Tampu \nametag{$^{1}$}} \Email{iulian.emil.tampu@liu.se}\\
\Name{Neda Haj-Hosseini \nametag{$^{1}$}} \Email{neda.haj.hosseini@liu.se}\\
\Name{Anders Eklund \nametag{$^{1,2,3}$}} \Email{anders.eklund@liu.se}\\
\addr $^{1}$  Department of Biomedical Engineering, Link\"{o}ping University, Sweden\\
\addr $^{2}$ Department of Computer and Information Science, Link\"{o}ping University, Sweden\\
\addr $^{3}$ Center for Medical Image Science and Visualization, Link\"{o}ping University, Sweden\\
}

\begin{document}

\maketitle

\begin{abstract}
Effective, robust, and automatic tools for~brain tumor segmentation are needed for~the~extraction of~information useful in~treatment planning. Recently, convolutional neural networks have shown remarkable performance in~the~identification of~tumor regions in~magnetic resonance (MR) images. Context-aware artificial intelligence is an~emerging concept for~the~development of~deep learning applications for~computer-aided medical image analysis. A large portion of~the~current research is devoted to the~development of~new network architectures to improve segmentation accuracy by~using context-aware mechanisms. In~this work, it is investigated whether or not the~addition of~contextual information from the~brain anatomy in~the~form of~white matter (WM), gray matter (GM), and cerebrospinal fluid (CSF) masks and probability maps improves U-Net-based brain tumor segmentation. The~BraTS2020 dataset was used to train and test two standard 3D U-Net (nnU-Net) models that, in~addition to the~conventional MR image modalities, used the~anatomical contextual information as~extra channels in~the~form of~binary masks (CIM) or~probability maps (CIP). For~comparison, a baseline model (BLM) that only used the~conventional MR image modalities was also trained. The~impact of~adding contextual information was investigated in~terms of~overall segmentation accuracy, model training time, domain generalization, and compensation for~fewer MR modalities available for~each subject. Median (mean) Dice scores of~90.2 (81.9), 90.2 (81.9), and 90.0 (82.1) were obtained on~the~official BraTS2020 validation dataset (125 subjects) for~BLM, CIM, and CIP, respectively. Results show that there is no statistically significant difference when comparing Dice scores between the~baseline model and the~contextual information models {(\emph{p}~$>$~0.05)}, even when comparing performances for~high and low grade tumors independently. In~a few low grade cases where improvement was seen, the~number of~false positives was reduced. Moreover, no improvements were found when considering model training time or~domain generalization. Only in~the~case of~compensation for~fewer MR modalities available for~each subject did the~addition of~anatomical contextual information significantly improve {(\emph{p}~<~0.05)} the~segmentation of~the~whole tumor. In~conclusion, there is no overall significant improvement in~segmentation performance when using anatomical contextual information in~the~form of~either binary WM, GM, and CSF masks or~probability maps as~extra~channels.
\end{abstract}


\section{Introduction}
Generally, patients diagnosed with brain tumor undergo radical treatment which can include a combination of~surgical tumor resection, radiotherapy, and chemotherapy~\cite{davis2016glioblastoma}. In~case of~surgery, a major factor that influences patient survival and postoperative morbidity is the~extent of~the~resection~\cite{davis2016glioblastoma, d2017extent}. Treatment planning depends extensively on~diagnostic radiology images for~the~identification of~the~tumor, key information for~balancing the~extent of~the~treatment target with the~collateral effects. 

MR imaging modalities, such as~T1-weighted (T1w), T1-weighted with post-contrast gadolinium enhancement (T1Gd), T2-weighted (T2w), and T2 fluid attenuated inversion recovery (FLAIR), are commonly used for~the~identification of~the~tumor~\cite{juratli2019radiographic}. Reliable tools for~the~extraction of~relevant information from the~MR images are needed. For~this, manual annotation of~brain tumors is commonly practiced in~clinical routine~\cite{visser2019inter}; however, this is a~time consuming and labor-intensive task. Moreover, manual annotation is not objective, with poor agreement between specialists~\cite{kubben2010intraobserver}. Automatic methods could overcome these limitations, providing a faster and objective identification of~the~tumor sub-regions. 

Automatic segmentation of~brain tumor structures in~MR images is challenging and has attracted a great research interest. Among the~proposed methods~\cite{tiwari2020brain}, convolutional neural networks (CNNs) have shown state-of-the-art performance, ranking first in~the~multimodal Brain Tumor Image Segmentation Benchmark (BraTS) challenge during recent years~\cite{bakas2018identifying}. Given the~automatic feature extraction of~CNNs~\cite{icsin2016review}, the~majority of~the~research is focused on~designing network architectures which provide better accuracy for~the~segmentation task. One of~the~most popular CNN architectures is U-Net~\cite{ronneberger2015u}, which introduced skip connections between the~layers in~the~network. A plethora of~U-Net-like architectures have since then been developed, including, among others, laborious skip connections strategies~\cite{zhou2018unet++} and attention mechanisms~\cite{noori2019attention}. 
However, Isensee et al.~\cite{isensee2018no}, who obtained top performance using a well-trained U-Net, showed that improving segmentation performance is not only a matter of~adjusting the~network architecture. The~choice of~loss function, training strategy, augmentation, and post-processing showed to have a large impact on~the~segmentation performance.

An emerging topic in~artificial intelligence applications, including computer-aided medical interventions, is context-awareness that will allow algorithms to use the~information from the~surrounding and perform segmentation of~images considering the~anatomy context~\cite{vercauteren2019cai4cai} and, thus, potentially improve the~outcome. The~latest literature describes different approaches for~context-aware deep learning including auto-context strategies, changing the~model architecture, and/or providing additional contextual information during training. Examples of~auto-context strategies used to improve model segmentation performance can be seen in Reference~\cite{tu2009auto, liu2020iouc}. In~particular, Reference~\cite{liu2020iouc} implemented auto-context in~their 3D symmetric fully convolutional neural network by~combining multi modal MR images with 3D Haar features with the~purpose of~improving brain tumor segmentation.

A number of~attempts have been made to evaluate the~impact of~the~introduction of~context-aware blocks in~the~model architecture on~brain tumor segmentation~\cite{ahmad2020context, chandra2018context, liu2020canet, pei2020context, le2020multi}. For~example, Pei et al.~\cite{pei2020context} used a context-aware deep neural network which thanks to~a~context encoding module between the~encoder and the~decoder part of~the~network, helped in~overcoming the~class imbalance problem that challenges brain tumor segmentation. However, such an~implementation does not allow a comparison between the~model accuracies with and without the~context encoding module; thus, the~contribution of~context information cannot be assessed.

Another approach for~achieving context-awareness is to provide the~network with more information~\cite{wachinger2018deepnat, shen2017boundary, shen2017efficient, kao2018brain}. Wachinger et al.~\cite{wachinger2018deepnat} included brain spectral coordinates information while training a patch-based deep CNN for~brain anatomy segmentation. The~authors argued that patches lose context information when propagating through the~network, with possible confusion due to the~symmetry of~the~brain. During training, each patch was supplemented with location information obtained from a spectral-based parameterization of~the~brain volume. Interestingly, the~additional information was provided intermediately in~the~network, concatenating the~context information to the~feature maps of~the~initial convolutional layers. In~two studies, Shen et al.~\cite{shen2017boundary, shen2017efficient} instead added four channels to~the~input of~their fully convolutional network in~addition to the~four conventional MR modalities. The~additional information consisted of~symmetry maps computed on~all the~MR modalities, describing the~asymmetry between the~brain hemispheres introduced by~the~tumor. Kao et al.~\cite{kao2018brain} included brain parcellation information during the~training of~a 3D U-Net as~means of~location information. In~their work, the~authors registered the~MNI152 1~mm brain parcellation atlas~\cite{desikan2006automated} to each BraTS subject, obtaining a mapping of~each voxel into one of~the~defined 69 brain structures. 

The aim of~this study is to expand this line of~thought and investigate if using anatomical contextual information as~additional input channels can improve brain tumor segmentation performance considering four aspects: (1) segmentation accuracy when training on~multimodal MR images, (2) model training time, (3) compensation for~fewer MR modalities available for~each subject, and (4) domain generalization. All four aspects are studied also with respect to low grade glioma (LGG) and high grade glioma (HGG) cases independently. Anatomical contextual information is defined in~this study as~white matter (WM), gray matter (GM), and cerebrospinal fluid (CSF) masks or~probability maps obtained automatically using an~automatic segmentation tool.

\section{Materials and Methods}
\subsection{Dataset}
Contextual information in~the~form of~binary WM, GM, and CSF masks and probability maps was obtained using FMRIB’s automated segmentation tool (FAST)~\cite{zhang2001segmentation} applied on~the~T1w MR volumes, each with normalized and zero-centered intensity. The~difference between the~FAST masks obtained from the~raw T1w and the~intensity normalized and zero-centered T1w volumes was minor. Of the~total 369 subjects, 92\% showed less than 10\% difference in~voxel classification (WM, GM, or~CSF). The~intensity normalized and zero-centered volumes were used instead of~the~raw data, since a preliminary investigation of~the~proposed method indicated that segmentation quality was lower when using contextual information from raw T1w data compared to when it was obtained from the~intensity normalized and zero-centered volumes. As Tudorascu et al. described~\cite{tudorascu2016reproducibility}, methods that use spatial priors during the~brain anatomy segmentation, such as~the~methods in~the~Statistical Parametric Mapping (SPM)~\cite{ashburner2014spm12} or~FreeSurfer~\cite{bruce2012freesurfer} softwares, may perform poorly on~diseased brains that contain deformations. Brain tumors can induce substantial deformation of~the~brain structures, making the~intensity-based FAST tool more suitable for~obtaining the~contextual information used in~this study, even if it is not specificately designed for~patients with tumors. An initial qualitative investigation on~obtaining the~anatomical segmentation through SPM showed that WM, GM, and CSF masks and probability maps lacked detail and were distorted. Given the~qualitatively assessed higher quality of~the~soft tissue masks, and that no spatial priors are used during the~anatomical segmentation, FAST was used in~this study.

\subsection{Anatomical Contextual Information}
Contextual information in~the~form of~binary WM, GM, and CSF masks and probability maps was obtained using FMRIB’s automated segmentation tool (FAST)~\cite{zhang2001segmentation} applied on~the~T1w MR volumes, each with normalized and zero-centered intensity. The~difference between the~FAST masks obtained from the~raw T1w and the~intensity normalized and zero-centered T1w volumes was minor. Of the~total 369 subjects, 92\% showed less than 10\% difference in~voxel classification (WM, GM, or~CSF). The~intensity normalized and zero-centered volumes were used instead of~the~raw data, since a preliminary investigation of~the~proposed method indicated that segmentation quality was lower when using contextual information from raw T1w data compared to when it was obtained from the~intensity normalized and zero-centered volumes. As Tudorascu et al. described~\cite{tudorascu2016reproducibility}, methods that use spatial priors during the~brain anatomy segmentation, such as~the~methods in~the~Statistical Parametric Mapping (SPM)~\cite{ashburner2014spm12} or~FreeSurfer~\cite{bruce2012freesurfer} softwares, may perform poorly on~diseased brains that contain deformations. Brain tumors
can induce substantial deformation of~the~brain structures, making the~intensity-based FAST tool more suitable for~obtaining the~contextual information used in~this study, even if it is not specificately designed for~patients with tumors. An initial qualitative investigation on~obtaining the~anatomical segmentation through SPM showed that WM, GM, and CSF masks and probability maps lacked detail and were distorted. Given the~qualitatively assessed higher quality of~the~soft tissue masks, and that no spatial priors are used during the~anatomical segmentation, FAST was used in~this study.

\begin{figure}[!t]
\floatconts
  {fig:method}
  {\caption{Schematic of~the~proposed method showing the~segmentation for~one subject when using all four conventional MR modalities (T1w~=~T1-weighted, T1Gd~=~T1-weighted with post-contrast gadolinium enhancement, T2w~=~T2-weighted, and FLAIR~=~T2 fluid attenuated inversion recovery) and the~3D U-Net-based deep learning model. By applying FMRIB's automated segmentation tool (FAST) to the~intensity normalized and zero-centered T1w volumes, the~contextual information is obtained as~white matter (WM), gray matter (GM), and cerebrospinal fluid (CSF) masks. The~red patch in~the~contextual masks shows where the~tumor is located. In~this region, the~contextual masks are distorted. The~final segmentation, obtained as~an~ensemble of~three cross-validation (CV) folds and provides regions of~tumor core (TC), enhancing tumor (ET), and edema (ED), shown here in~green, red, and blue, respectively. BLM is the~baseline model, and CIM and CIP are the~contextual information models using binary masks and probability maps, respectively.}}
  {\includegraphics[width=1\textwidth]{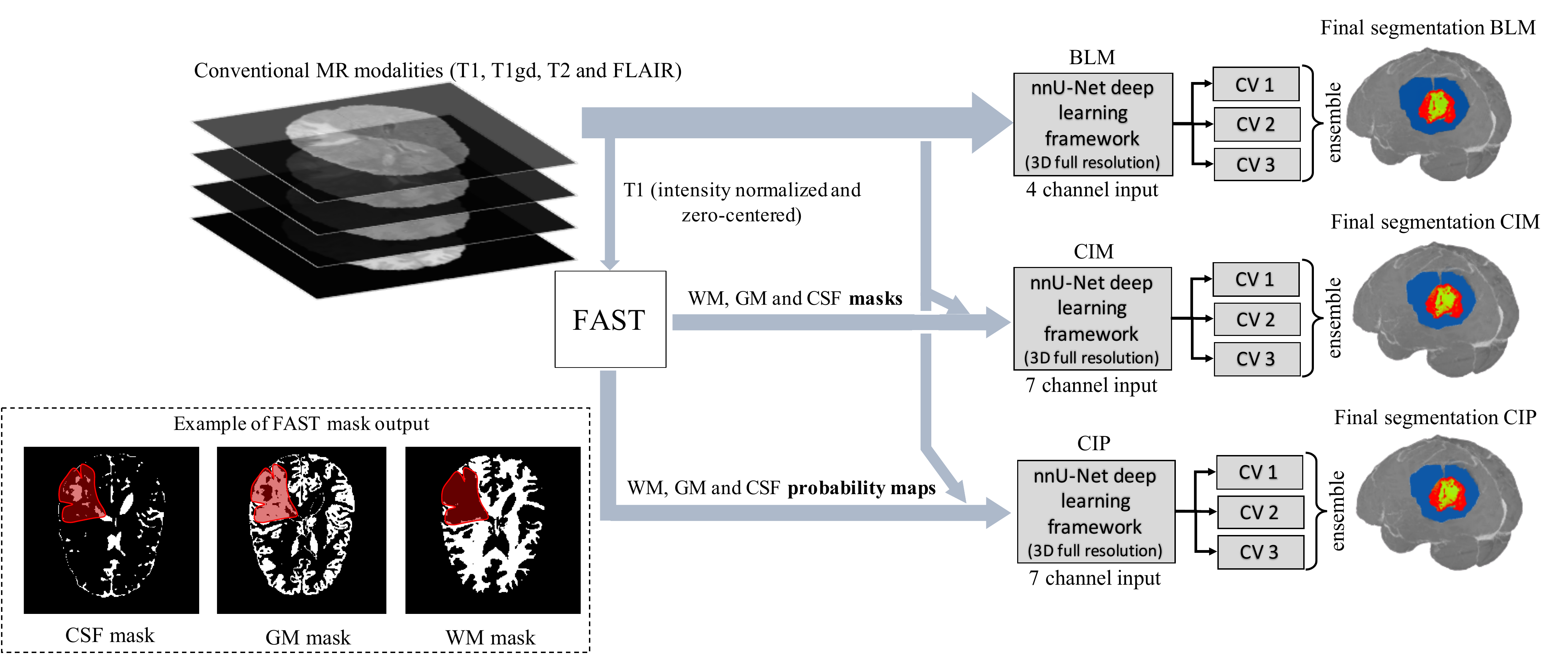}}
\end{figure}

\subsection{Model}
\figureref{fig:method} shows an~overview of~the~methodology, where the~nnU-Net deep learning framework~\cite{isensee2021nnu} was used. There are two reasons for~this choice: (1) nnU-Net’s repeated success in~the~BraTS challenges in~the~recent years shows the~reliability of~this framework, which could be difficult to achieve with an~in-house model, and (2) this allows reproducibility of~the~presented investigation. nnU-Net is built upon the~3D U-Net architecture and automatically the~tunes network hyperparameters based on~the~training dataset and hardware available. Among others, the~framework tunes the~number of~convolutional layers, input patch size and batch size~\cite{isensee2021nnu}. In~this study, the~3D full resolution U-Net configuration was adopted, and four NVIDIA Tesla V100 GPUs (32 GB memory) were used for~the~training. During training, the~sum of~Dice and cross-entropy loss was minimized using stochastic gradient descent with Nesterov momentum ($\mu=0.99$). The~number of~training epochs was automatically set to 1000 by~nnU-Net, without any early stopping strategies. Each epoch consisted of~250 mini-batches.

\subsection{Evaluation and Statistical Methods}
To investigate if the~addition of~contextual information has an~impact on~glioma segmentation performance, three models were trained that differed in~the~use or~not of~the~contextual information: a baseline model (BLM) with input channels chosen among the~four conventional MR modalities provided by~BraTS, and two contextual information models both with three additional channels compared to BLM to accommodate the~extra information obtained from FAST. One contextual information model used binary WM, GM, and CSF masks (CIM), while the~other model used the~WM, GM, and CSF probability maps (CIP). {A 3-fold} cross validation scheme was used to train each setting described below. After training, the~segmentation of~the~test subjects was obtained as~an~ensemble of~the~predictions of~the~three models trained through cross validation. Dice score~\cite{menze2014multimodal} and 95\% Hausdorff distance (HD)~\cite{menze2014multimodal} on~the~segmentation targets were obtained through the~automatic validation system, on~both the~official BraTS2020 validation dataset (125 cases) and an~independent test dataset (36 cases), described in \mbox{Section \ref{sec:Multimodal}}. The~non-parametric Wilcoxon signed-rank test was used to test the~null hypothesis of~no difference between the~baseline model and the~contextual information models, at a significance level of~5\%. Statistical analysis was performed in~IBM\textsuperscript{\tiny\textregistered} SPSS\textsuperscript{\tiny\textregistered} (Version 27.0, Armonk,~NY,~USA,~IBM~Corp).

\subsection{Multimodal MR Model Training}\label{sec:Multimodal}
To study the~effect of~anatomical contextual information on~segmentation performance, all four conventional MR modalities were used as~input to the~three models, with CIM and CIP additionally using the~anatomical contextual information, as~described above. In~addition to the~official BraTS2020 validation dataset, 36 subjects containing an~equal number of~HGGs and LGGs were randomly selected from the~training dataset as~the~independent test dataset, with the~remaining 333 subjects used for~training. The~choice of~an~independent test dataset with control over the~tumor grades allows us to investigate the~effect that anatomical contextual information has on~the~segmentation of~LGGs and HGGs independently. Moreover, to understand the~impact of~contextual information on~the~model training time, the~validation loss curves saved by~nnU-Net were analyzed a~posteriori for~these models. Training was considered finished when the~validation loss did not improve over 50 epochs.

\subsection{Compensation for~Fewer MR Modalities}
To explore if anatomical contextual information could compensate for~the~missing information when only one MR modality is used as~input, the~three models were trained and tested, similarly as~in~Section~\ref{sec:Multimodal}, with only T1Gd provided instead of~four MR images per subject. T1Gd was selected, among the~other MR modalities, given that (1) it provides contrast of~the~tumor core region compared to the~surrounding healthy tissue~\cite{bakas2018identifying} and (2) because T1w is already used by~FAST to obtain the~anatomical contextual information.

\subsection{Domain Generalization}
Finally, to investigate if the~addition of~contextual information improves domain generalization, the~three models were trained on~BraTS cases from a single institute and tested on~data from a variety of~institutes. In~particular, a total of~69 (35~LGGs and 34~HGGs), among the~369 cases, were identified to originate from one of~the~19 institutes that contributed to the~BraTS2020 dataset. Identification of~the~institutes was possible using the~information from Reference~\cite{bakas2018identifying, bakas2017advancing} and the~BraTS name mapping information. Models were trained using all conventional MR modalities, and the~69 cases were excluded from the~independent test dataset. 

\section{Results}

\begin{figure}[t!]
\centering
\includegraphics[width=13cm]{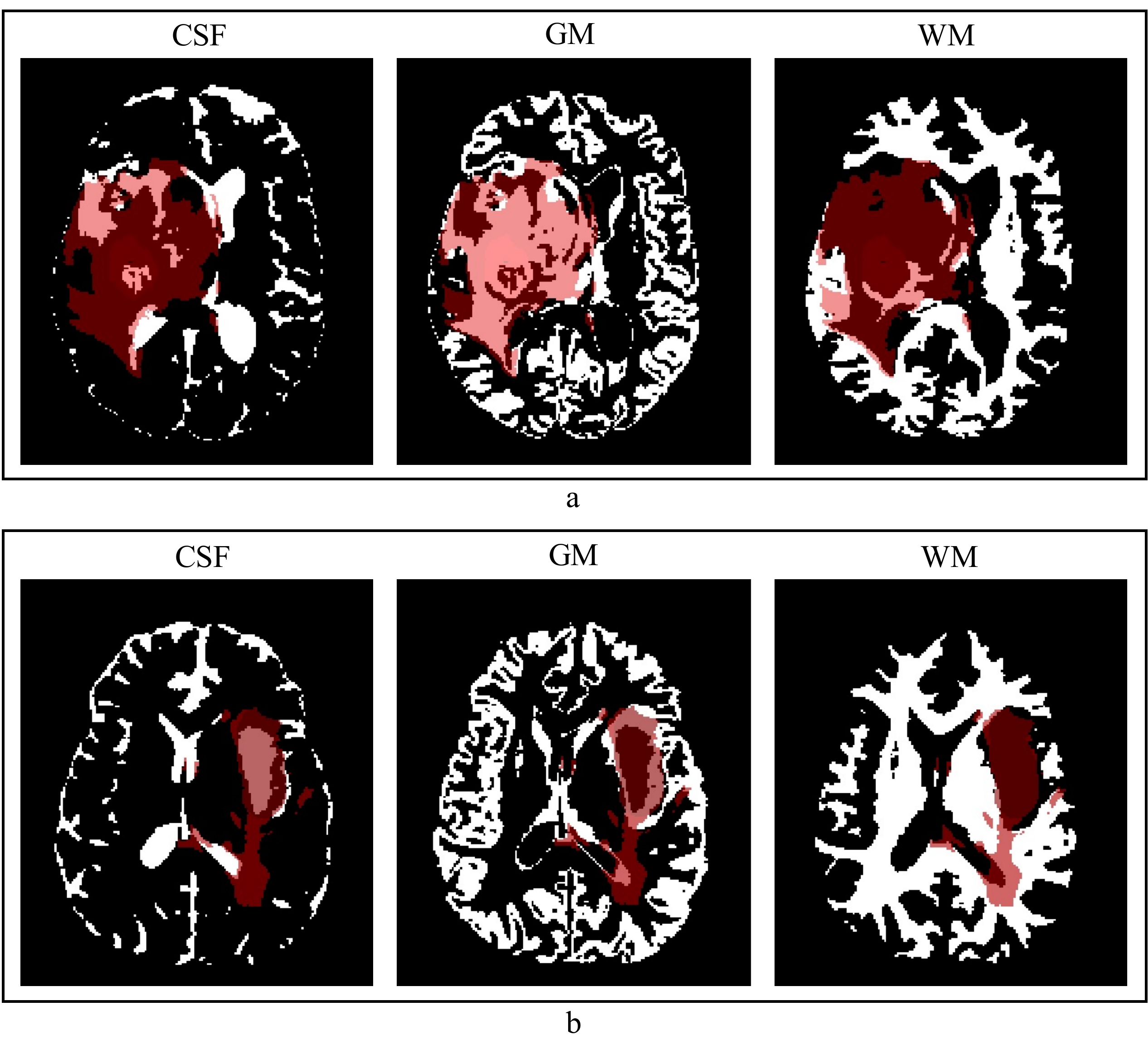}
\caption{Examples of~contextual information obtained using FMRIB's automated segmentation tool are shown for~two cases in~an~axial slice with the~tumor annotation overlay in~red. From left to right, cerebrospinal fluid (CSF), gray matter (GM), and white matter (WM) masks. (\textbf{a}) A case where the~soft tissue masks are highly distorted; (\textbf{b}) a case where masks are only distorted in~the~tumor region.}
\label{fig:contextual_info_example}
\end{figure}

An example of~contextual information obtained using FAST can be seen in \figureref{fig:contextual_info_example}, where the~cross-section of~CSF, GM, and WM masks are shown for~two subjects. By visually inspecting the~FAST results, the~soft tissue segmentations are descriptive of~the~brain WM, GM, and CSF structures, with the~masks and probability maps being distorted only in~the~regions where the~tumor is located or~proximal to it.

\subsection{Segmentation Accuracy for~Multimodal MR Model Training}
\begin{figure}[t!]
\centering
\includegraphics[width=13.5 cm]{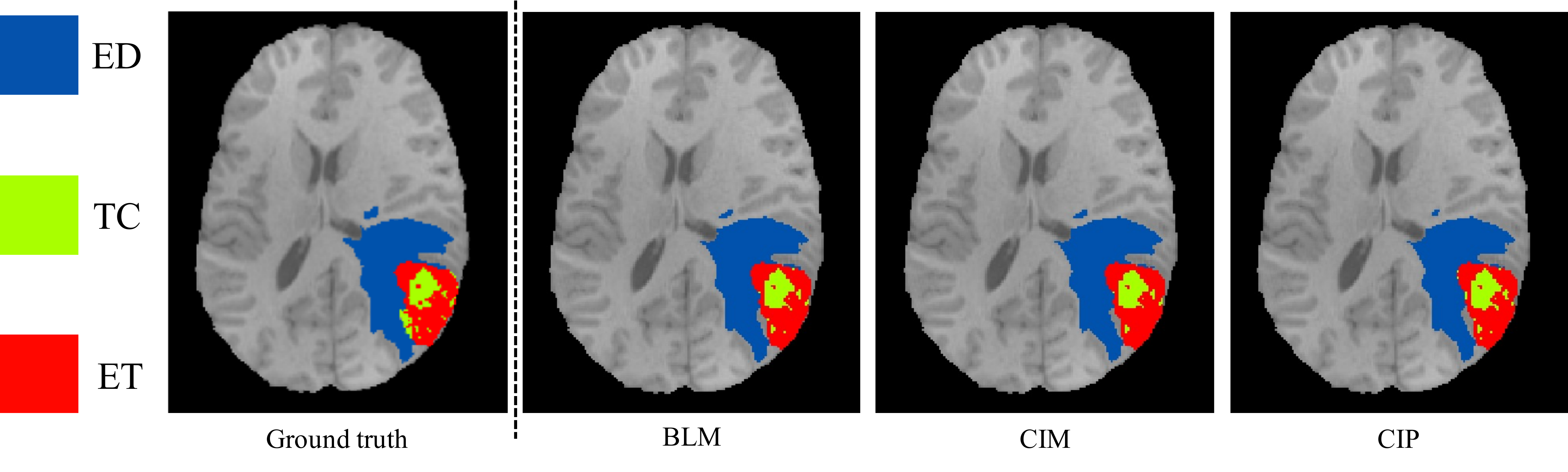}
\caption{Segmentation results obtained for~one of~the~independent test subjects and for~the~three models (BLM~=~baseline model, CIM~=~contextual information model using binary masks, and CIP using probability maps) are shown as~colored labels overlaid on~a T1-weighted axial slice. Tumor core (TC), enhancing tumor (ET), and edema (ED) are shown in~green, red, and blue, respectively. Dice scores for~BLM, CIM, and CIP for~the~presented test case are 92.54, 92.53, and 92.33, respectively.}\label{fig:results_segmentation_example}
\end{figure}

\begin{figure}[t!]
\centering
\includegraphics[width=13.5 cm]{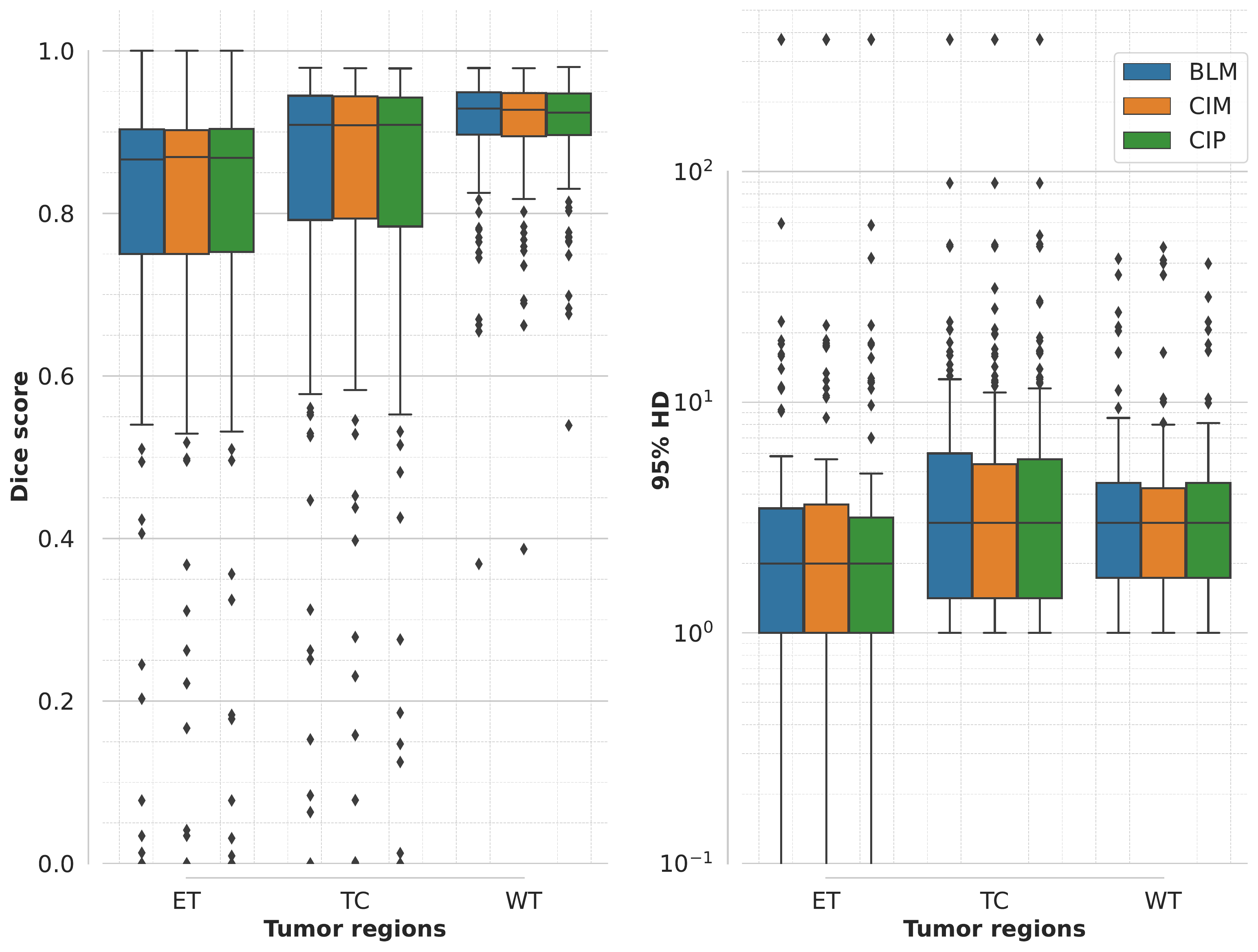}
\caption{Boxplot of~Dice scores and 95\% Hausdorff distance (HD) of~different models trained on~multimodal MR images computed for~the~official BraTS2020 validation dataset (125 subjects). Boxplots show median and range (with box showing 25\% and 75\% quantiles) for~each model (BLM~=~baseline model, CIM~=~contextual information model using binary masks, and CIP using probability maps) and segmentation target (ET~=~enhancing tumor, TC~=~tumor core, and WT~=~whole tumor). The~statistical analysis showed no significant difference (\emph{p}~$>$~0.05) when comparing CIM and CIP with BLM.}\label{fig:result_multiModality}
\end{figure}

An example of~segmentation results in~an~axial slice for~one of~the~independent test samples for~BLM, CIM, and CIP models when using all the~available MR modalities is shown \mbox{in~Figure~\ref{fig:results_segmentation_example}}. Performance of~the~three models on~the~official validation dataset (125~subjects) is shown as~boxplots in~Figure~\ref{fig:result_multiModality}, where Dice scores and 95\% HD are reported for~all segmentation targets. Dice scores’ median values (mean) across target regions were 90.15 (81.85), 90.17 (81.87), and 90.04 (82.06), for~BLM, CIM, and CIP, respectively. \mbox{Table~\ref{tab:LGGvsHGG_multiModality}} summarizes the~median Dice scores and and 95\% HD obtained on~the~independent test dataset for~LGG and HGG cases separately, showing that HGGs are overall better segmented than LGGs. When comparing CIM and CIP to BLM across the~different tumor regions, no statistically significant difference (\emph{p}~$>$~0.05) in~Dice scores was found when analyzing the~results from both the~independent test dataset and the~official BraTS2020 validation set. Moreover, no significant difference (\emph{p}~$>$~0.05) was observed when comparing the~effect of~contextual information in~segmenting LGG and HGG cases separately, with HGG showing slightly lower \emph{p}-values. When looking at the~cases that showed at least 5\% improvement in~mean Dice score when using anatomical contextual information, it could be seen that the~enhancing tumor region was better segmented. Among the~subjects in~the~independent test dataset, all of~those with improved mean Dice score (5.6\% of~the~total subjects) were LGGs, with the~contextual information models avoiding false positives for~the~enhancing tumor region. Table~\ref{tab:model_comparisons} summarizes mean Dice scores across tumor regions for~studies that implemented context-awareness by~means of~auto-context, architectural changes, or~additional contextual information. The~mean Dice scores obtained in~this study are in~the~same range as~results previously reported in~literature. Note that the~disparity in~segmentation performance of~the~result presented here compared to Reference~\cite{isensee2021nnu} is due to the~fact that, in~this work, only the~3D full resolution U-Net configuration was used, instead of~a combination of~2D U-Net, 3D low resolution, and 3D full resolution U-Net, that can be trained and ensembled using the~nnU-Net framework, at the~cost of~a longer training time.

\begin{table}[ht!]
\floatconts
  {tab:LGGvsHGG_multiModality}%
  {\caption{Median Dice scores and 95\% Hausdorff distance (HD) across target regions (ET~=~enhancing tumor, TC~=~tumor core, and WT~=~whole tumor) for~the~different models (BLM~=~baseline model, CIM~=~contextual information model using binary masks, and CIP using probability maps) trained on~multimodal MR images. Results are shown for~low grade glioma (LGG) and high grade glioma (HGG) cases independently computed on~the~independent dataset (36 subjects). There was no significant difference (\emph{p}~$>$~0.05) when comparing CIM and CIP to BLM for~neither of~LGG and HGG cases. The~results in~bold correspond to the~best performing model in~each tissue region.}}
{\makebox[\linewidth]{
\begin{tabular}{c c c c c c c c }
\hline
\multirow{3}{*}{\textbf{Grade}} & \multirow{3}{*}{\textbf{Model}} & \multicolumn{3}{c}{\textbf{Median Dice Score}} & \multicolumn{3}{c}{\textbf{Median 95\% HD}}\\
& & \multicolumn{3}{c}{\textbf{[min, max]}} & \multicolumn{3}{c}{\textbf{[min, max]}}\\ 
& & \textbf{TC} & \textbf{ET }& \textbf{WT} & \textbf{TC} & \textbf{ET} & \textbf{WT} \\
\hline
\multirow{6}{*}{\begin{tabular}[x]{@{}c@{}}LGG\\ \footnotesize{(18 cases)} \end{tabular}} & \multirow{2}{*}{BLM} & 79.96 & 36.72 & 93.28 & 7.21 & 39.66 & \textbf{3.00} \\
& & \small{[0.00, 98.39]} & \small{[0.00, 100.00]} & \small{[0.00, 97.09]} & \small{[1.00, 373.13]} & \small{[0.00, 373.13]} & \small{[1.00, 373.13]}\\\cline{2-8}
& \multirow{2}{*}{CIM}& 81.95 & 44.87 & \textbf{93.51} & \textbf{6.98} & \textbf{25.18} & \textbf{3.00} \\
& & \small{[0.00, 98.38]} & \small{[0.00, 100.00]} & \small{[0.00, 96.98]} & \small{[1.00, 373.13]} & \small{[0.00, 373.13]} & \small{[1.41, 373.13]}\\\cline{2-8}
& \multirow{2}{*}{CIP}& \textbf{82.27} & \textbf{51.27} & 93.22 & 7.64 & 23.17 & 3.08\\
& & \small{[0.00, 80.00]} & \small{[0.00, 100.00]} & \small{[10.08, 97.18]} & \small{[1.00, 373.13]} & \small{[0.00, 373.13]} & \small{[1.00, 18.49]}\\\cline{1-8}
\multirow{6}{*}{\begin{tabular}[x]{@{}c@{}}HGG\\ \footnotesize{(18 cases)}\end{tabular}} & \multirow{2}{*}{BLM} & \textbf{94.83} & \textbf{89.22} & 92.79 & \textbf{1.41} & \textbf{1.41} & 2.00\\
& & \small{[65.76, 97.56]} & \small{[75.54, 96.57]} & \small{[86.64, 97.19]} & \small{[1.00, 21.47]} & \small{[1.00, 3.00]} & \small{[1.00, 7.87]}\\\cline{2-8}
& \multirow{2}{*}{CIM}& 94.80 & 89.16 & \textbf{92.94} & 1.57 & \textbf{1.41} & \textbf{1.87}\\
& & \small{[64.22, 97.37]} & \small{[74.74, 96.17]} & \small{[86.90, 97.17]} & \small{[1.00, 22.03]} & \small{[1.00, 3.00]} & \small{[1.00, 7.87]}\\\cline{2-8}
& \multirow{2}{*}{CIP}& 94.54 & 89.12 & 92.74 & 1.57 & \textbf{1.41} & \textbf{1.87}\\
& & \small{[67.68, 97.48]} & \small{[74.21, 96.55]} & \small{[86.96, 97.27]} & \small{[1.00, 14.87]} & \small{[1.00, 3.00]} & \small{[1.00, 7.55]}\\
\hline
\end{tabular}}}
\end{table}

\begin{table}[ht!]
\floatconts
  {tab:model_comparisons}%
  {\caption{Mean Dice scores reported by~similar studies using context-aware methods. $^*$ identifies the~method ranking first in~BraTS2020 (not using context-awareness mechanisms). $^1$ obtained via official BraTS test dataset, $^2$ obtained via official BraTS validation datasets, $^3$ obtained from a randomly selected subset of~BraTS training data. ET~=~enhancing tumor, TC~=~tumor core, and WT~=~whole tumor, and BLM~=~baseline model, CIM~=~contextual information model using binary masks, and CIP using probability maps. The~results in~bold correspond to the~best performing model in~each tissue region.}}
{\makebox[\linewidth]{
\begin{tabular}{c c c c c c}
\hline
\multirow{2}{*}{\textbf{Model}} & \multirow{2}{*}{\textbf{Dataset}} & \multicolumn{3}{c}{\textbf{Mean Dice Score}} &  \multirow{2}{*}{\begin{tabular}[x]{@{}c@{}}\footnotesize{\textbf{Claimed Improvement}}\\  \footnotesize{\textbf{Using Context Awareness}}\end{tabular}}\\
 & & \textbf{TC} & \textbf{ET} & \textbf{WT} \\
\hline
Isensee et al.~\cite{isensee2020nnu} $^*$ & \footnotesize{BraTS2020 $^1$} & 85.95 & 82.03 & 88.95 & \footnotesize{no context-aware mechanism used}\\
\hline
Liu et al. \cite{liu2020iouc} & \footnotesize{BraTS2017 $^3$} & 84.00 & 78.00 & 89.00 & \multirow{5}{*}{\begin{tabular}[x]{@{}c@{}}comparison between\\ models with and without \\ context-aware mechanism \\ not available\end{tabular}}\\
Liu et al. \cite{liu2020canet} & \footnotesize{BraTS2019 $^2$} & 85.10 & 75.90 & 88.50 & \\
Ahmad et al. \cite{ahmad2020context} & \footnotesize{BraTS2020 $^1$} & 84.67 & 79.10 & 89.12 & \\
Chandra et al. \cite{chandra2018context} & \footnotesize{BraTS2018 $^1$} & 73.33 & 61.82 & 82.99 &\\
Pei et al. \cite{pei2020context} &\footnotesize{BraTS2019/20 $^3$} & 83.50 & \textbf{82.10} & 89.50 &\\
\hline
Shen et al. \cite{shen2017efficient} & \footnotesize{BraTS2013 $^3$} & 71.80 & 72.50 & 88.70 & 2--3\% (no \emph{p}-value)\\
Shen et al. \cite{shen2017boundary} & \footnotesize{BraTS2015 $^1$} & 82.00 & 75.00 & 87.00 & 1.3\% (\emph{p}-value $<$ 0.01) \\
Kao et al. \cite{kao2018brain} & \footnotesize{BraTS2018 $^1$} & 79.30	& 74.90 & 87.50 & 1--2\% (no \emph{p}-value)\\
Le et al. \cite{le2020multi} & \footnotesize{BraTS2018 $^2$} & \textbf{88.88} & 81.41 & \textbf{90.95} & 2\% (no \emph{p}-value)\\
\hline
BLM & \multirow{3}{*}{\begin{tabular}[x]{@{}c@{}}\footnotesize{BraTS2020 $^3$}\\(\footnotesize{36 cases)}\end{tabular}} & 81.80 & 67.20 & 90.80 & \multirow{3}{*}{\begin{tabular}[x]{@{}c@{}}none\end{tabular}} \\
CIM & & 81.90 & 77.00 & 90.10 \\
CIP & & 81.80 & 70.40 & 90.50 \\
\hline
BLM & \multirow{3}{*}{\begin{tabular}[x]{@{}c@{}}\footnotesize{BraTS2020 $^2$}\\\footnotesize{(125 cases)}\end{tabular}} & 81.60 & 73.41 & 90.54 & \multirow{3}{*}{\begin{tabular}[x]{@{}c@{}}none\end{tabular}} \\
CIM & & 81.61 & 73.43 & 90.58 \\
CIP & & 81.44 & 74.05 & 90.69 \\
\hline
\end{tabular}}}
\end{table}

\subsection{Model Training Time for~Multimodal MR Model Training}
From the~a posteriori analysis of~the~validation loss curves, the~baseline model trained 12 and 5 h (46 and 9 epochs) faster than CIM and CIP, respectively, when looking at the~average values across the~three folds. Average training times and epochs for~the~three models are summarized in~Table~\ref{tab:training_times}.

\begin{table}[ht]
\floatconts
  {tab:training_times}%
  {\caption{Average training epochs and time across folds for~the~different models (BLM~=~baseline model, CIM~=~contextual information model using binary masks, and CIP using probability maps.}}
{\makebox[\linewidth]{
\begin{tabular}{c c c c c c}
\hline
\textbf{Model }&\textbf{ Average Training Time [hh:mm:ss] }&\textbf{ Epochs [250 Mini-Batches Each]} \\
         \hline
         BLM & 8:15:11 & 79\\
         CIM & 20:02:39 & 140\\
         CIP & 12:58:57 & 103 \\
\hline
\end{tabular}}}
\end{table}

\subsection{Compensation for~Fewer MR Modalities}
\begin{figure}[!h]
\centering
\includegraphics[width=12.7cm]{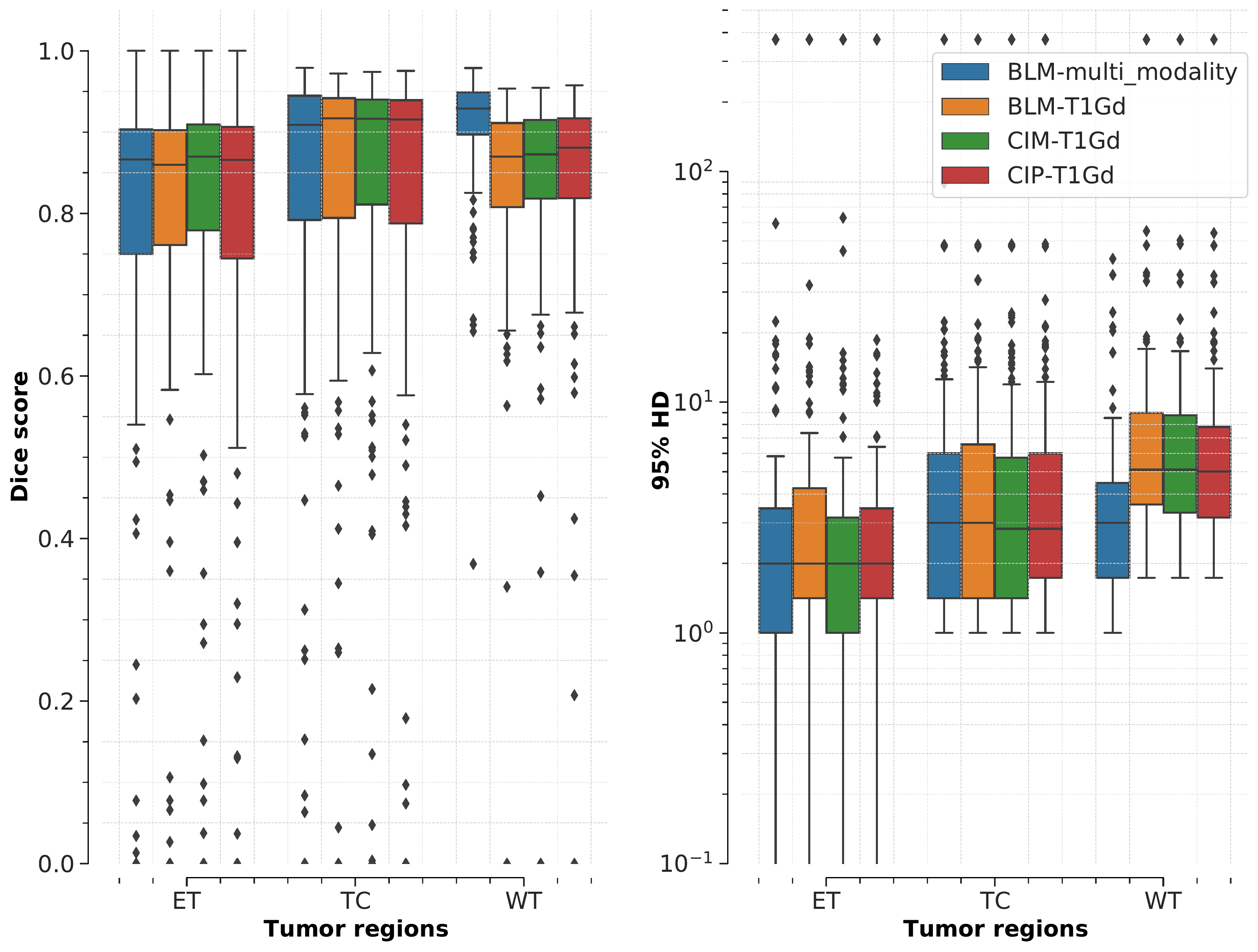}
\caption{Boxplot of~Dice scores and 95\% Hausdorff distance (HD) of~different models (BLM~=~baseline model, CIM~=~contextual information model using binary masks, and CIP using probability maps) trained using only T1-weighted with post-gadolinium enhancement (T1Gd) as~single MR modality computed for~the~official BraTS2020 validation dataset (125 subjects). For~comparison, the~performance of~BLM trained on~data from all institutes is presented (blue box). Boxplots show median and range (with box showing 25\% and 75\% quantiles) for~each model and segmentation target (ET~=~enhancing tumor, TC~=~tumor core, and WT~=~whole tumor). By comparing CIM and CIP with the~BLM, Dice scores of~the~WT region show a statistically significant difference (\emph{p}~$<$~0.05).}\label{fig:limited_MR_modalities}
\end{figure}

Segmentation performance results on~the~BraTS2020 validation set (125 subjects) for~BLM, CIM, and CIP when trained using only T1Gd as~MR modality are summarized in~Figure~\ref{fig:limited_MR_modalities}. Dice score values for~TC and ET regions are similar when compared to the~models trained on~all the~four MR modalities. The~whole tumor region segmentation, on~the~other hand, shows a decrease in~performance that can be attributed to the~lack of~the~contrast in~the~T1Gd between edema region and surrounding tissue, that is present in~FLAIR.

 When comparing the~models trained only on~T1Gd on~the~BraTS validation dataset, the~Dice score for~the~whole tumor region is significantly improved (\emph{p}~$<$~0.05) for~both contextual information models compared to the~baseline model (also after Bonferroni correction for~multiple comparisons). Considering the~results for~LGG and HGG cases separately computed on~the~independent test dataset (36 subjects), no significant difference could be found between the~models, not even with respect to the~whole tumor region. Median Dice and 95\% HD for~the~independent test set are summarized in~\mbox{Table~\ref{tab:LGGvsHGG_limited_MR_modalities}}.

\begin{table}[ht!]
\floatconts
  {tab:LGGvsHGG_limited_MR_modalities}%
  {\caption{Median Dice scores and 95\% Hausdorff distance (HD) across target regions (ET~=~enhancing tumor, TC~=~tumor core, and WT~=~whole tumor) for~the~different models (BLM~=~baseline model, CIM~=~contextual information model using binary masks, and CIP using probability maps) trained using only T1-weighted with post-gadolinium enhancement as~MR modality. Values are shown for~low grade glioma (LGG) and high grade glioma (HGG) cases from the~independent test dataset. There was no significant difference (\emph{p}~$>$~0.05) for~neither of~LGG and HGG cases when comparing CIM and CIP with BLM. The~results in~bold correspond to the~best performing model in~each tissue region.}}
{\makebox[\linewidth]{
\begin{tabular}{c c c c c c c c}
\hline
\multirow{3}{*}{\textbf{Grade}} & \multirow{3}{*}{\textbf{Model}} & \multicolumn{3}{c}{\textbf{Median Dice Score}} & \multicolumn{3}{c}{\textbf{Median 95\% HD}}\\
& & \multicolumn{3}{c}{\textbf{[min, max]}} & \multicolumn{3}{c}{\textbf{[min, max]}}\\
& & \textbf{TC} & \textbf{ET} & \textbf{WT} & \textbf{TC} & \textbf{ET} & \textbf{WT} \\
\hline
\multirow{6}{*}{\begin{tabular}[x]{@{}c@{}}LGG\\ \footnotesize{(18 cases)} \end{tabular}} & \multirow{2}{*}{BLM} & \textbf{79.99} & \textbf{61.08} & \textbf{86.65} & 9.25 & 176.67 & 86.01 \\
& & \small{[0.00, 98.06]} & \small{[0.00, 100.00]} & \small{[0.00, 95,34]} & \small{[1.00, 373.13]} & \small{[0.00, 373.13]} & \small{[0.00, 373.13]}\\\cline{2-8}
& \multirow{2}{*}{CIM}& 78.21 & 37.93 & 83.85 & \textbf{6.42} & 37.85 & \textbf{9.26}\\
& & \small{[0.00, 98.04]} & \small{[0.00, 100.00]} & \small{[0.00, 95.16]} & \small{[1.00, 373.13]} & \small{[0.00, 373.13]} & \small{[2.00, 373.13]}\\\cline{2-8}
& \multirow{2}{*}{CIP}& 79.58 & 52.22 & 83.27 & 7.87 & \textbf{23.43} & 10.37\\
& & \small{[0.00, 98.21]} & \small{[0.00, 100.00]} & \small{[0.00, 94.87]} & \small{[1.00, 373.13]} & \small{[0.00, 373.13]} & \small{[2.24, 373.13]}\\\hline
\multirow{6}{*}{\begin{tabular}[x]{@{}c@{}}HGG\\ \footnotesize{(18 cases)}\end{tabular}} & \multirow{2}{*}{BLM} & \textbf{94.41} & 89.70 & \textbf{90.35} & \textbf{1.87} & \textbf{1.41} & \textbf{3.86}\\
& & \small{[87.55, 97.10]} & \small{[69.71, 96.29]} & \small{[71.90, 94.51]} & \small{[1.00, 4.90]} & \small{[1.00, 2.83]} & \small{[1.73, 15.17]}\\\cline{2-8}
& \multirow{2}{*}{CIM}& 93.90 & \textbf{89.86} & 89.58 & \textbf{1.87} & \textbf{1.41} & 3.93\\
& & \small{[66.94, 97.06]} & \small{[70.93, 96.41]} & \small{[77.27, 94.49]} & \small{[1.00, 13.49]} & \small{[1.00, 3.00]} & \small{[2.00, 13.00]}\\\cline{2-8}
& \multirow{2}{*}{CIP}& 94.10 & 89.75 & 89.96 & \textbf{1.87} & \textbf{1.41} & \textbf{3.86}\\
& & \small{[63.08, 97.34]} & \small{[72.14, 96.45]} & \small{[79.54, 94.61]} & \small{[1.00, 22.38]} & \small{[1.00, 3.00]} & \small{[1.73, 10.72]}\\
\hline
\end{tabular}}}
\end{table}

\subsection{Domain Generalization}
\begin{figure}[!h]
\centering
\includegraphics[width=12.5cm]{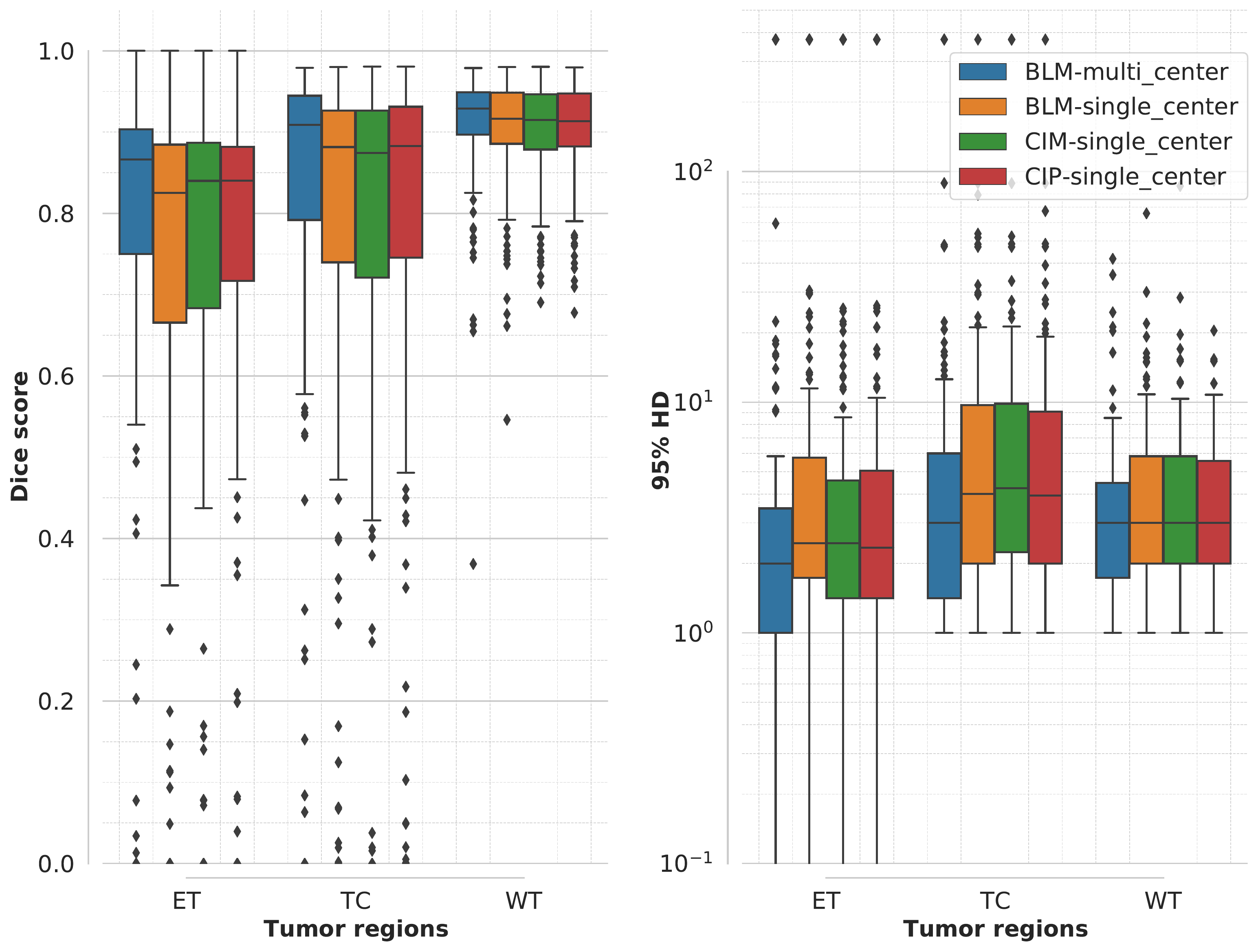}
\caption{Boxplot of~Dice scores and 95\% Hausdorff distance (HD) for~the~different models (BLM~=~baseline model, CIM~=~contextual information model using binary masks, and CIP using probability maps) trained on~single center data computed for~BraTS2020 validation dataset. For~comparison, the~performance of~BLM trained on~data from all institutes is presented (blue box). Boxplots show median and range (with box showing 25\% and 75\% quantiles) for~each model and segmentation target (ET~=~enhancing tumor, TC~=~tumor core, and WT~=~whole tumor). The~statistical difference was insignificant (\emph{p}~$>$~0.05) when comparing CIM and CIP with BLM.}\label{fig:result_singleCenter}
\end{figure}

Dice scores and 95\% HD values on~the~official BraTS validation dataset obtained for~the~models trained on~data from a single institute are summarized in~Figure~\ref{fig:result_singleCenter}. Compared to the~baseline model trained on~data from all the~institutes, performance is lower especially for~the~TC and ET tumor regions. The~drop in~performance of~all the~models trained only on~single-center data shows the~impact of~domain shift between the~training and test datasets. The~addition of~anatomical contextual information does not help in~this aspect, since the~three models trained on~single-center data have similar performances, which are all significantly lower (\emph{p}~$<$~0.05) than the~one of~the~model trained on~all the~MR data available. Results on~the~independent test dataset, summarized in~Table~\ref{tab:LGGvsHGG_singleCenter}, show a similar trend for~both LGG and HGG cases. 

\begin{table}[ht!]
\floatconts
  {tab:LGGvsHGG_singleCenter}%
  {\caption{Median Dice scores and 95\% Hausdorff distance (HD) across target regions (ET~=~enhancing tumor, TC~=~tumor core, and WT~=~whole tumor) for~the~different models (BLM~=~baseline model, CIM~=~contextual information model using binary masks, and CIP using probability maps) trained on~single center data. Values are shown for~low grade glioma (LGG) and high grade glioma (HGG) cases from the~independent test dataset (36 subjects). The~statistical difference was insignificant (\emph{p}~$>$~0.05) for~either LGG or~HGG cases when comparing CIM and CIP with BLM. In~bold, the~results of~the~better performing model for~each segmentation target.}}
{\makebox[\linewidth]{
\begin{tabular}{c c c c c c c c}
\hline
\multirow{3}{*}{\textbf{Grade}} & \multirow{3}{*}{\textbf{Model}} & \multicolumn{3}{c}{\textbf{Median Dice Score}} & \multicolumn{3}{c}{\textbf{Median 95\% HD}}\\
& & \multicolumn{3}{c}{\textbf{[min, max]}} & \multicolumn{3}{c}{\textbf{[min, max]}}\\
& & \textbf{TC} & \textbf{ET} & \textbf{WT} & \textbf{TC} & \textbf{ET} & \textbf{WT} \\
\hline
\multirow{3}{*}{\textbf{Grade}} & \multirow{3}{*}{\textbf{Model}} & \multicolumn{3}{c}{\textbf{Median Dice Score}} & \multicolumn{3}{c}{\textbf{Median 95\% HD}}\\
& & \multicolumn{3}{c}{\textbf{[min, max]}} & \multicolumn{3}{c}{\textbf{[min, max]}}\\
& & \textbf{TC} & \textbf{ET} & \textbf{WT} & \textbf{TC} & \textbf{ET} & \textbf{WT} \\
\hline
\multirow{6}{*}{\begin{tabular}[x]{@{}c@{}}LGG\\ \footnotesize{(18 cases)} \end{tabular}} & \multirow{2}{*}{BLM} & \textbf{78.47} & \textbf{60.81} & 92.77 & \textbf{6.40} & \textbf{7.64} & 2.73\\
& & \small{[0.00, 97.31]} & \small{[0.00, 100.00]} & \small{[76.48, 97.11]} & \small{[0.00, 97.31]} & \small{[1.00, 373.13]} & \small{[0.00, 373.13]} \\\cline{2-8}
& \multirow{2}{*}{CIM}& 75.99 & 56.71 & 93.02 & 6.71 & 13.42 & 2.53\\
& & \small{[0.00, 96.85]} & \small{[0.00, 100.00]} & \small{[84.68, 97.21]} & \small{[1.00, 373.13]} & \small{[0.00, 373.13]} & \small{[1.41, 12.69]}\\\cline{2-8}
& \multirow{2}{*}{CIP}& 73.53 & 60.19 & \textbf{93.29} &6.54 & 13.98 & \textbf{2.45}\\
& & \small{[0.00, 97.05]} & \small{[0.00, 100.00]} & \small{[14.81, 96.94]} & \small{[1.00, 373.13]} & \small{[0.00, 373.13]} & \small{[1.00, 14.78]}\\\hline
\multirow{6}{*}{\begin{tabular}[x]{@{}c@{}}HGG\\ \footnotesize{(18 cases)}\end{tabular}} & \multirow{2}{*}{BLM}& 91.68 & 85.66 & 89.30 & \textbf{2.24} & 2.00 & \textbf{3.23} \\
& & \small{[63.67, 96.15]} & \small{[53.57, 94.62]} & \small{[76.57, 96,62]} & \small{[1.00, 12.37]} & \small{[1.00, 6.63]} & \small{[1.00, 75.21]}\\\cline{2-8}
& \multirow{2}{*}{CIM}& 90.97 & \textbf{86.37} & 89.13 & 2.34 & 2.00 & 3.67\\
& & \small{[67.18, 96.62]} & \small{[53.79, 94.29]} & \small{[80.83, 96.84]} & \small{[1.14, 10.05]} & \small{[1.00, 6.48]} & \small{[1.00, 13.64]}\\\cline{2-8}
& \multirow{2}{*}{CIP}& \textbf{92.16} & 86.33 & \textbf{89.41} & 2.27 & \textbf{1.23} &3.38 \\
& & \small{[63.55, 96.36]} & \small{[54.59, 94.25]} & \small{[81.16, 96.63]} & \small{[1.41, 13.19]} & \small{[1.00, 6.40]} & \small{[1.00, 10.20]}\\
\hline
\end{tabular}}}
\end{table}

\newpage
\section{Discussion}

The effect of~anatomical contextual information on~brain tumor segmentation was investigated with respect to segmentation performance, model training, model generalization, and compensation for~fewer MR modalities.

\subsection*{Segmentation Accuracy and Training Time for~Multimodal MR Model Training}
Glioma segmentation performance in~this study showed no significant improvement when comparing models trained with the~addition of~anatomical contextual information as~input channels along with the~conventional MR modalities. A possible reason for~the~observed results may be found in~how the~WM, GM, and CSF information are computed. FAST uses pixel intensity and spatial information for~the~segmentation. Arguably, this is very similar to what a U-Net architecture is using when trained for~semantic segmentation. Thus, it is possible that the~network is independently creating a representation of~WM, GM, and CSF at some stage during training from the~conventional MR modalities, nullifying the~additional information. However, providing such information already as~input channels did not speed up model training, given that BLM trained faster compared to the~contextual information models based on~the~a posteriori analysis of~the~validation loss curves. The~addition of~the~contextual information does not improve segmentation performance but instead increases the~model convergence time, suggesting that the~extra information is not used and makes the~segmentation problem harder to solve.

Direct comparison between the~obtained results with other works is partially possible, given the~differences in~testing datasets. The~results presented here are in~the~same range of~reported findings of~studies that used additional contextual information during model training. Overall, the~reported results in~literature and the~ones obtained in~this study show that the~inclusion of~context-awareness, by~means of~model architecture changes or~additional information as~input to the~network, has~marginal or~no improvement on~glioma segmentation~\cite{liu2020iouc, shen2017boundary, shen2017efficient, kao2018brain, ahmad2020context, chandra2018context, liu2020canet, pei2020context, le2020multi}. This should not discourage future research on~the~topic, but instead promote studies that exploit contextual information for~brain tumor segmentation by~other approaches and perhaps a combination of~the~currently implemented methods, {i.e.,} context-aware blocks and additional contextual information as~input to the~network. 

\subsection*{Quality of~the~Anatomical Contextual Information}
Another reason for~why the~model does not use the~additional anatomical information may be found in~the~quality of~the~WM, GM, and CSF binary masks and probability maps. As shown in~Figure~\ref{fig:contextual_info_example}, the~WM, GM, and CSF masks are distorted in~the~brain region containing the~tumor, which may not help the~network. One possible way of~investigating this aspect is to compare model performance with anatomical contextual information obtained automatically or~from manual annotations. However, the~amount of~time that would be needed for~the~annotation of~WM, GM, and CSF of~each subject is exceedingly large making this comparison unfeasible. Another possible approach is to obtain the~anatomical information from quantitative MRI (qMRI)~\cite{tofts2005quantitative}. By quantitatively measuring the~relaxation times of~tissues in~the~brain, qMRI can provide probability maps for~WM, GM, and CSF. In~contrast to the~automatically generated probability maps used in~this study, the~ones obtained through qMRI are not derived quantities; thus, the~information given in~each of~the~input channels is unique and is not a different version of~the~same information. This does increase the~amount of~information that the~model can actually use for~the~segmentation task. Given that the~BraTS2020 dataset does not provide qMRI data, this approach remains open for~investigation.

\subsection*{Compensation for~Fewer MR Modalities}
Reducing the~MR modalities needed to be acquired could have a positive impact on~patient hospitalization experience and on~healthcare economy, since a shorter time would be needed for~the~patient to be in~the~MR scanner and more patients could be scanned~\cite{hollingsworth2015reducing}. For~this reason, here, it was investigated whether or not the~addition of~anatomical information could compensate for~the~decrease in~segmentation performance caused by~using only one MR modality (T1Gd) as~input to the~model. Results show that only the~segmentation of~the~whole tumor region is affected by~the~lack of~the~excluded MR modalities. This is not surprising since the~WT includes the~ED region, which is not visible in~T1Gd. However, the~addition of~contextual information marginally improves WT segmentation, suggesting that the~WM, GM, and CSF masks help the~model to better identify the~edema region.

\subsection*{Domain Generalization}
Domain shift is a challenge that today's deep learning models in~general have to address when intended for~real world applications~\cite{zhou2021domain}. In~the~context of~medical image segmentation, models trained on~data from a single center or~scanner often struggle to retain segmentation performance when tested or~used on~data that originates from different centers or~scanners. As described by~Zhou et al.~\cite{zhou2021domain}, domain generalization is still a challenge that hinders the~expansion of~deep learning methods for~real world applications. Anatomical contextual information has shown no impact on~domain generalization, given that the~models trained on~single center data and with WM, GM, and CSF as~extra input channels have suffered from a similar drop in~the~performance compared to the~baseline performance as~the~model trained without it. A possible reason for~this, as~discussed above, is that the~model is not using the~additional anatomical information.

\subsection*{Effect of~Contextual Information on~LGG and HGG Cases}
Overall, anatomical contextual information does not improve segmentation performance when considering high and low grade cases separately. Even if no statistically significant improvement could be found between the~baseline and the~contextual information models, in~some of~the~LGG cases, the~addition of~contextual information reduced the~number of~false positives for~the~enhancing tumor region. The~low number of~LGGs in~the~training dataset, compared to HGGs, could bias the~model in~always segmenting an~enhancing tumor region. However, in~LGG cases, this tumor sub-region is missing, which leads to a high number of~false positives when the~model segments it. When using anatomical contextual information, the~model could better discriminate HGG and LGG cases, thus avoiding segmenting the~enhancing tumor region that is not present in~LGGs. Yet, the~statistical comparison does not support this hypothesis, since no statistically significant difference was found between the~baseline model and the~contextual information models for~the~cases in~this study. Nevertheless, a higher number of~LGG test samples are needed to study this effect.

\subsection*{Future Perspectives}
A possible approach to investigate is providing contextual information not as~additional channels to the~input, but after the~initial convolutional layers. Wachinger~et~al.~\cite{wachinger2018deepnat} showed improved neuroanatomy segmentation performance when concatenating context information intermediately in~the~convolutional network. Moreover, the~authors also showed that different types of~context information affect the~performance differently; a combination of~spectral and cartesian-based parametrization of~the~brain volume yielded a better performance than when only one of~the~two was used, suggesting that they might contain complementary information. Thus, investigation could focus on~finding different types of~contextual information and their combinations. Future research could also address if using additional MR modalities ({e.g.,} diffusion MR imaging) would improve brain tumor segmentation. The~major factor hindering such investigation at the~current moment is the~lack of~a standardized open access dataset which includes extra MR modalities. 

\section{Conclusions}
Anatomical contextual information in~the~form of~binary WM, GM, and CSF and probability maps was obtained in~this study using the~automatic FAST segmentation tool. The~addition of~anatomical contextual information as~extra channels to the~network shows no statistically significant difference in~tumor segmentation performance when using a standardized 3D U-Net architecture and conventional multimodal MR images as~dataset. Only in~the~case of~using one conventional MR modality (T1Gd) did the~addition of~the~anatomical contextual information show to significantly improve whole tumor region segmentation. No statistically significant improvements could be seen when investigating HGG and LGG cases separately, nor~when considering model training time, domain generalization, and compensation for~fewer MR modalities. Overall, context-aware approaches implemented for~brain tumor segmentation in~the~recent literature show only minor or~no improvements. This suggests that effective integration of~context awareness in~deep learning models for~glioma segmentation has yet to be explored.

\newpage
\bibliography{main}

\end{document}